\documentclass[twocolumn]{aastex631}

\usepackage{newtxtext,newtxmath}

\usepackage[T1]{fontenc}
\usepackage{ae,aecompl}

\usepackage{graphicx}	
\usepackage{amsmath}	
\usepackage{amssymb}	

\usepackage[normalem]{ulem}
\usepackage{ragged2e}
\usepackage{xcolor}
\usepackage{color}
\usepackage{booktabs}
\usepackage{comment}
\usepackage{float}
\usepackage{wrapfig}

\begin{document}





\title{Formation of a nuclear star cluster through the inspiral of globular clusters: a case study of the dwarf elliptical galaxy UGC 7346}

\correspondingauthor{Fazeel Mahmood Khan}
\email{fmk5060@nyu.edu}

\author{Ismael Khan}
\affiliation{Department of Space Science, Institute of Space Technology (IST), Islamabad 44000, Pakistan}
\affiliation{Space and Astrophysics Research Lab (SARL), National Centre of GIS and Space Applications (NCGSA), Islamabad 44000, Pakistan}

\author[0000-0002-5707-4268]{Fazeel Mahmood Khan}
\affiliation{New York University Abu Dhabi, PO Box 129188, Abu Dhabi, United Arab Emirates}
\affiliation{Center for Astrophysics and Space Science (CASS), New York University Abu Dhabi}
\affiliation{Space and Astrophysics Research Lab (SARL), National Centre of GIS and Space Applications (NCGSA), Islamabad 44000, Pakistan}

\author[0000-0002-8171-6507]{Andrea V. Macci\`o}
\affiliation{New York University Abu Dhabi, PO Box 129188, Abu Dhabi, United Arab Emirates}
\affiliation{Center for Astrophysics and Space Science (CASS), New York University Abu Dhabi}

\author[0000-0003-2227-1322]{Kelly~Holley-Bockelmann}
\affiliation{ Department of Physics and Astronomy, Vanderbilt University, Nashville, TN 37240, USA}
\affiliation{Department of Physics, Fisk University, Nashville, TN 37208, USA}

\author[0000-0003-4176-152X]{Peter~Berczik}
\affiliation{Nicolaus Copernicus Astronomical Centre, Polish Academy of Sciences, ul. Bartycka 18, 00-716 Warsaw, Poland}
\affiliation{Main Astronomical Observatory, National Academy of Sciences of Ukraine, 27 Akademika Zabolotnoho St, 03143 Kyiv, Ukraine}


\begin{abstract}

 Nuclear star clusters (NSCs) are dense stellar environments located in the center of most galaxies. NSCs are thought to form through two primary methods; through the inspiral of globular clusters (GCs) to the galactic center due to dynamical friction, and through in-situ star formation. Recent observations of dwarf elliptical galaxy UGC 7346 suggest that it might be undergoing NSC formation due to the presence of multiple GCs near its photometric center. We perform direct N-body simulations of nine GCs belonging to UGC 7346's GC system to investigate whether their eventual infall to the galactic center would result in the formation of a NSC. 
Our simulations indicate that GCs inspiral leads to the formation of a central stellar over-density relative to the background profile of the host galaxy within $\sim$1.5 Gyr, corresponding to a nuclear star cluster with a typical mass of $(4.1$–$4.5)\times10^5 ~M_{\odot}$. Several key structural parameters of the newly formed NSC, including the S\'ersic index, effective radius, and central stellar density, lie well within the range observed for NSCs.
We also test a hypothetical scenario in which some of the infalling GCs have larger masses ($M \sim 10^6 M_{\odot}$), resulting in the formation of a more massive NSC whose mass and size are more consistent with observations. Our results suggest that inspiral of GCs is a viable channel for assembling a significant mass in the shape of NSC in the center of dwarf galaxies and that UGC 7346 will host a NSC at its center in about 2-3 Gyr. 
\end{abstract}

\keywords{Nuclear star clusters -- galaxies: dwarf galaxies, galactic nuclei -- methods: numerical}

	

\section{Introduction}\label{sec:intro}

The central part of dwarf and late-type galaxies, including the Milky Way, often demonstrate "excessive light" relative to inward extrapolation of the galaxy's light profile. This excess light is attributed to a dense, compact stellar system known as a nuclear star cluster (NSC), which typically has a mass of $10^{5} - 10^{8} M_{\odot}$ and a half-light radius of only a few parsecs \citep{neumayer_etal_2020}. Unlike GCs, NSCs can host both metal-rich young and metal-poor old stellar populations \citep{kacharov_etal_2018, fahrion2020a}. The dominant population varies with host galaxy type and mass \citep{neumayer_etal_2020}, providing key constraints on NSC formation histories \citep{fahrion_etal_2021}.

Nuclear star clusters (NSCs) are generally thought to form through two primary channels:
\begin{enumerate}
    \item Infall of globular clusters (GCs) toward the galactic center driven by dynamical friction \citep{tremaine1975formation}. This mechanism is often invoked to explain the presence of old, metal-poor stars in NSCs.
    \item In-situ star formation in the galactic nucleus, fuelled by gas and dust funneled inward \citep{loose1982bursts}. This mechanism naturally accounts for the young, metal-rich stellar populations observed in many NSCs.
\end{enumerate}


The current understanding of the NSCs formation scenario is that both channels work in conjunction \citep{antonini_etal_2015}, however, which channel provides the greatest contribution depends on the morphology of the host galaxy in question and its GC system \citep{fahrion_etal_2021,fahrion2022}.
\citet{neumayer_etal_2020} proposed that the primary NSC formation mechanism shifts from GC collapse to in-situ star formation at a host galaxy mass of roughly $10^9$ $M_{\odot}$, since that is also where the occupation fraction peaks. \citet{fahrion_etal_2021} tested this claim by performing a detailed study of the ages, metallicities, and star formation histories of nucleated galaxies present in the Fornax cluster. Their results suggest that low mass NSCs (found in low mass galaxies) are significantly metal poorer than their host galaxy, and also possess predominantly old stars (inactive star formation). Higher mass NSCs (located in higher mass galaxies) on the other hand are consistently metal rich than their host galaxy, and possess rich star formation histories. This difference in properties between the host galaxy and its NSC point toward the dominant NSC formation mechanism, being GC collapse for low-mass nucleated galaxies (hence the poor metallicities and old ages), and in-situ star formation for high-mass nucleated galaxies. Galaxies of intermediate mass (around $10^9\, M_{\odot}$) show signs of both mechanisms in action.

The NSC occupation fraction exhibits a clear dependence on the host galaxy’s mass: it increases with galaxy mass, reaches a maximum around $10^9$ $M_{\odot}$, and then declines at higher masses
\citep{ordenes_briceno_2018_ngfs_iv,sanchez_janssen_2019_ngvs_xxiii}.
For low-mass dwarf galaxies the NSC occupation fraction is small, likely because several physical processes work against forming and retaining a compact nuclear component. Their shallow gravitational potential is inefficient at funneling gas into the very central regions and, even when gas does reach the center, stellar feedback can readily expel it, preventing the buildup of a long-lived, dense reservoir for repeated nuclear star-formation episodes. At the same time, the dynamical-friction timescales for GCs to spiral in from kiloparsec scales are very long in such low-mass systems, potentially leaving GCs stalled \citep{arca2017lack, kaur2018stalling}. While the tidal-disruption mass-loss rate scales with host-galaxy mass \citep{Moreno2024}, these stalled clusters would still undergo gradual mass loss and would therefore contribute only minimally to the central mass budget, even if they ultimately reached the center.

For host galaxy masses greater than $10^9$ $M_{\odot}$, the occupation fraction decreases, probably because such galaxies are also known to possess a central massive black hole (MBH). The tidal forces of the MBH disrupt any further growth of the NSC, and may even grow itself by accreting mass from NSC \citep{cote_etal_2006_acsvcs_viii}. More importantly, mergers of MBHs can disrupt NSCs through ejection of stars via gravitational slingshot mechanism \citep{Khan_Holley-Bockelmann2021}. \par
The occupation fraction of NSCs also depends on the environment in which the host galaxy resides (although still debated), specifically whether it is a field galaxy or present in a galactic cluster. The occupation fraction of early-type galaxies is enhanced if it is present in a galactic cluster \citep{ferguson_sandage_1989,lisker_etal_2007,paudel2025}, however observations of the \citet{georgiev_etal_2009} and \citet{georgiev_boker_2014} sample of late-type galaxies show that the occupation fraction remains relatively high regardless of whether it is present in a cluster or not.

\citet{antonini_etal_2012} performed N-body simulations to study the formation of NSC in the Milky Way. However, due to limited resolution the GCs were introduced at a distance of only 20 pc from the center. Furthermore, each GC was added sequentially—only after the previous one had reached the center and settled into an equilibrium distribution. The mass resolution in their simulations was also relatively low, of the order of 200 $M_{\odot}$ per particle. Despite these limitations, a reasonable model of the Milky Way NSC was produced.

\citet{arca_sedda_etal_2015_henize_2_10} modeled formation of NSC in galaxy Henize 2–10 through inspiral of GCs and concluded that such a channel is feasible in formation of NSC. The GCs initial positions were all within 200 pc and the mass resolution of GCs particle was chosen to be roughly 80 $M_{\odot}$. 

Here, we test the possible NSC formation scenario in recent observations of ongoing GC inspiral in dwarf galaxy UGC 7346 by performing detailed N-body simulations. The spatial extent of the GC system is much larger than previously studied cases (within 1 kpc), and the masses of GCs are of the order of $10^5\, M_{\odot}$ (compared to $10^6\, M_{\odot}$ in previous studies).
Additionally, we study the NSC formation scenario with a much better mass resolution of $25\, M_{\odot}$ per particle. We also explore the impact of initial conditions 
on the final outcome of evolution. \par 
Our manuscript is arranged as follows:
Section 2 describes N-body models of UGC 7346 galaxy and globular clusters invoked. We discuss the main results of our study in Section 3. Section 4 puts our findings in the context of existing literature, and finally Section 5 summarizes and concludes the paper. 



\begin{table*}[]
\centering
\begin{tabular}{@{}cccccccc@{}}
\toprule
ID &
  \begin{tabular}[c]{@{}c@{}}$r_{e_{1,2}},r_{e_3}$\\ ($pc$)\end{tabular} &
  \begin{tabular}[c]{@{}c@{}}$M_{1,2}, M_{3}$ \\ ($M_{\odot}$)\end{tabular} &
  $N_{1,2},N_3$ &
  \begin{tabular}[c]{@{}c@{}}$r_1,r_3$\\ ($pc$)\end{tabular} &
  \begin{tabular}[c]{@{}c@{}}$r_2$\\ ($pc$)\end{tabular} &
  \begin{tabular}[c]{@{}c@{}}$r_{t1},r_{t3}$\\ ($pc$)\end{tabular} &
  \begin{tabular}[c]{@{}c@{}}$r_{t2}$\\ ($pc$)\end{tabular} \\ \midrule
GC0 & 1.36,2.9 & $1.40 \times 10^5$,$1.40 \times 10^6$ & 5624,56241  & 532.1 & 653.8 & 62.5,134.7 & 66.4\\
GC1 & 1.36,2.9 & $2.26 \times 10^5$,$2.26 \times 10^6$ & 9079,90794  & 673.3 & 841.6 & 78.6,169.5 & 85.0\\
GC2 & 1.31,--- & $1.78 \times 10^5$,--- & 7145,---  & 289.6 & 845.4 & 58.8,--- & 78.6\\
GC3 & 2.03,2.9 & $1.34 \times 10^5$,$1.34 \times 10^6$ & 5371,53710  & 820.7 & 146.0 & 70.4,152.4 & 47.6\\
GC4 & 1.52,--- & $2.29 \times 10^5$,--- & 9163,---  & 162.2 & 480.6 & 57.8,--- & 71.6\\
GC5 & 1.58,--- & $2.96 \times 10^5$,--- & 11859,--- & 665.0 & 613.2 & 85.6,--- & 83.5\\
GC6 & 1.51,2.9 & $2.16 \times 10^5$,$2.16 \times 10^6$ & 8670,86707  & 706.3 & 840.8 & 78.7,169.6 & 83.7\\
GC7 & 1.51,2.9 & $1.73 \times 10^5$, $1.73 \times 10^6$ & 6951,69511  & 899.9 & 776.6 & 79.8,172.0 & 75.5\\
GC8 & 1.66,--- & $2.80 \times 10^5$,--- & 11221,--- & 449.9 & 653.0 & 75.3,--- & 83.6\\ \bottomrule
\end{tabular}
\caption{Table of parameters used for modeling the GCs in IC1, IC2 and IC3. Column 1: GC name. Column 2: Effective radius. Column 3: GC mass. Column 4: GC number of particles. Column 5: Starting galacto-centric distance in IC1 and IC3. Column 6: Starting galacto-centric distance in IC2. Column 7: Initial tidal radius in IC1 and IC3. Column 8: Initial tidal radius in IC2. }
\label{tab:gc-params}
\end{table*}

\section{Initial conditions}\label{sec:ics}

UGC 7346 is a dwarf elliptical galaxy located in the periphery of the Virgo Cluster of galaxies, with an estimated total stellar mass of $1.4 \times 10^9$ $M_{\odot}$ \citep{j_roman_etal_2023} (hereafter JR23). 
JR23 fitted UGC~7346 by a two-component model. The main component is extended ($R_e = 2.55$~kpc), and contains the bulk of the stellar mass
($1.2 \times 10^9\,M_\odot$) with a shallow Sérsic index ($n = 0.78$) and axes ratio ($b/a = 0.94$). The secondary component is more compact
($R_e = 0.88$~kpc), with similar Sérsic index and axes ratio ($n = 0.84$, $b/a = 0.91$), and contributes $2.4 \times 10^8\,M_\odot$.\par

JR23 suggest that UGC 7346 is going through the early stages of NSC formation based on its observed high central concentration of GCs.
They explain the high central concentration by arguing that the galaxy has experienced a past merger event, as evidenced by the existence of a stellar stream (see figure 1 of JR23), and the existence of an inner bluer stellar component \citep{benitez_llambay_etal_2016}. 
The merger event will have induced accelerated collapse of the GCs \citep{oh_lin_2000} which resulted in their current distribution.
Galaxy mergers can also lead to a central concentration of gas \citep{mihos1994ultraluminous}, which can contribute to building a NSC \citep{gray2025}. A notable example of this being NGC 6946, which through bar-driven gas infall has accumulated nearly $10^7\, M_{\odot}$ of gas within $30\, \text{pc}$ of its center \citep{schinnerer2007bar}.
\par

To model UGC 7346 and its GC system, we have used the AGAMA galaxy modeling architecture \citep{eugene_2019_agama}, which provides a complete set of tools to generate dynamical equilibrium models of galaxies.

\subsection{UGC 7346 galaxy model}\label{subsec:model_ugc7346}
In order to build an $N-$body model of UGC 7346 galaxy in dynamical equilibrium, we employed the de-projected S\'ersic profile \citep{prugniel_simien_1997,ter05}, given by:
\begin{equation}\label{eq:deprojected_sersic}
    \rho(R) = \rho_e \left( \frac{R}{R_e} \right)^{-p} exp \left[ -b_n \left( \frac{R}{R_e} \right)^{1/n} \right]
\end{equation}
Where $R_e$ is the effective radius, $n$ is the degree of central concentration (S\'ersic index), $\rho_e$ is the density at $R_e$, $b_n$ is a constant given by: 
\begin{equation}\label{eq:sersic_bn}
    b_n = 2n - \frac{1}{3} + \frac{0.009876}{n}
\end{equation}
and $p$ is a function of $n$, given as:
\begin{equation}\label{eq:PS_p}
    p = 1.0-\frac{0.6097}{n}+\frac{0.05463}{n^2}, \quad 0.6 <~ n <~ 10
\end{equation}
To reduce computational cost, we chose to model UGC 7346 as a single-component galaxy with a reduced total mass of $M_{gal}=5\times 10^8\, M_{\odot}$ instead of a two-component galaxy with main and secondary components. For our truncated model, we adopt a value of 0.84 for S\'ersic index $n$, and 1.056 kpc for the galaxy's effective radius. The N-body model is realized with 1 million particles and has intermediate to major axis ratio of $b/a = 0.9$ and minor to major axis ratio of $c/a = 0.8$. The mass distribution of our chosen model closely matches the original two-component model's distribution within 1 kpc, beyond which it is cut off deliberately. To model UGC 7346 and its GC system, we used the AGAMA galaxy modeling framework \citep{eugene_2019_agama}, which generates equilibrium galaxy models.

\subsection{Globular cluster models}\label{subsec:model_gcs}
JR23 identified a total of 14 GCs within $3R_e$ of the main component, of which nine are present within $1R_e$.
We chose to model the nine inner GCs to save computational cost and time.
Each GC was modeled using a generalized King model with $W_o=7$ \citep{king_1966, gieles_zocchi_2015, jordan_etal_2005_acsvcs_x}. 

We employed equation 12 of \cite{jordan_etal_2005_acsvcs_x} to calculate the half-light radii of our chosen GCs, given as:
\begin{equation}\label{eq:half_light_radius}
    r_h = r_{av}10^{(-0.016(\mu_z-21)+0.17[(g-z)_{gal}-1.5]+0.17[(g-z)_{gc}-1.2])}
\end{equation}
This equation returns the half-light radius for a GC with color $(g-z)_{gc}$, in a galaxy of local surface brightness $\mu_z$ and color $(g-z)_{gal}$. Here, $r_{av}=0.035\, \text{arcsec}$ as measured by \cite{jordan_etal_2005_acsvcs_x}. 

We estimate the mass of GCs from their $g$ band magnitude reported in JR23, assuming a mass-to-light ratio of 2 \citep{baumgardt_etal_2020}.
The calculated masses of chosen GCs are listed in table \ref{tab:gc-params}.

The total mass in GC system available for NSC formation is $1.8\times 10^6$ $M_{\odot}$. Finally, the initial tidal radii of each GC were calculated using the following equation:
\begin{equation}
    R_t = R\left( \frac{ M_{gc} }{ 2 M_{gal}(R) } \right) ^ {1/3}
\end{equation}
Here, $R$ denotes the current galacto-centric distance of the GC, $M_{\mathrm{gc}}$ is the GC mass at a given time step, and $M_{\mathrm{gal}}(R)$ is the galaxy mass enclosed within radius $R$.

\subsection{Numerical methods} \label{subsec:methods}
\begin{figure}
    \centering
    \includegraphics[width=\linewidth]{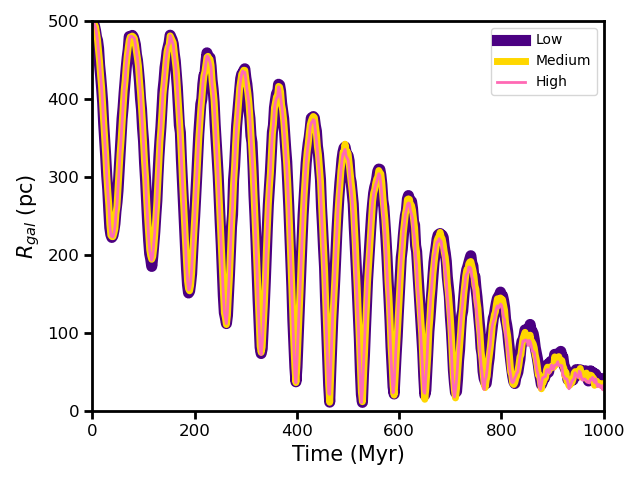}
    \centering
    \includegraphics[width=\linewidth]{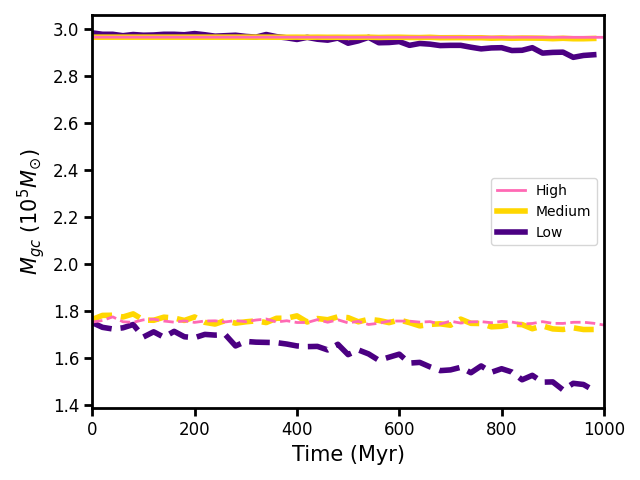}
    \caption{Top panel: Orbital decay of GC 5 in runs with low, intermediate and high resolution. In all cases the GCs reach the center almost at the same time irrespective of the mass resolution used in GC 5 model.
    Bottom panel: Mass loss of GC 5 in our test runs. Solid lines indicate mass within tidal radius and dashed lines indicate mass within $5\, R_e$.}
    \label{fig:test_ics_orbits}
\end{figure}

\begin{figure*}[p!]
    \centering
    \includegraphics[width=1.1\linewidth]{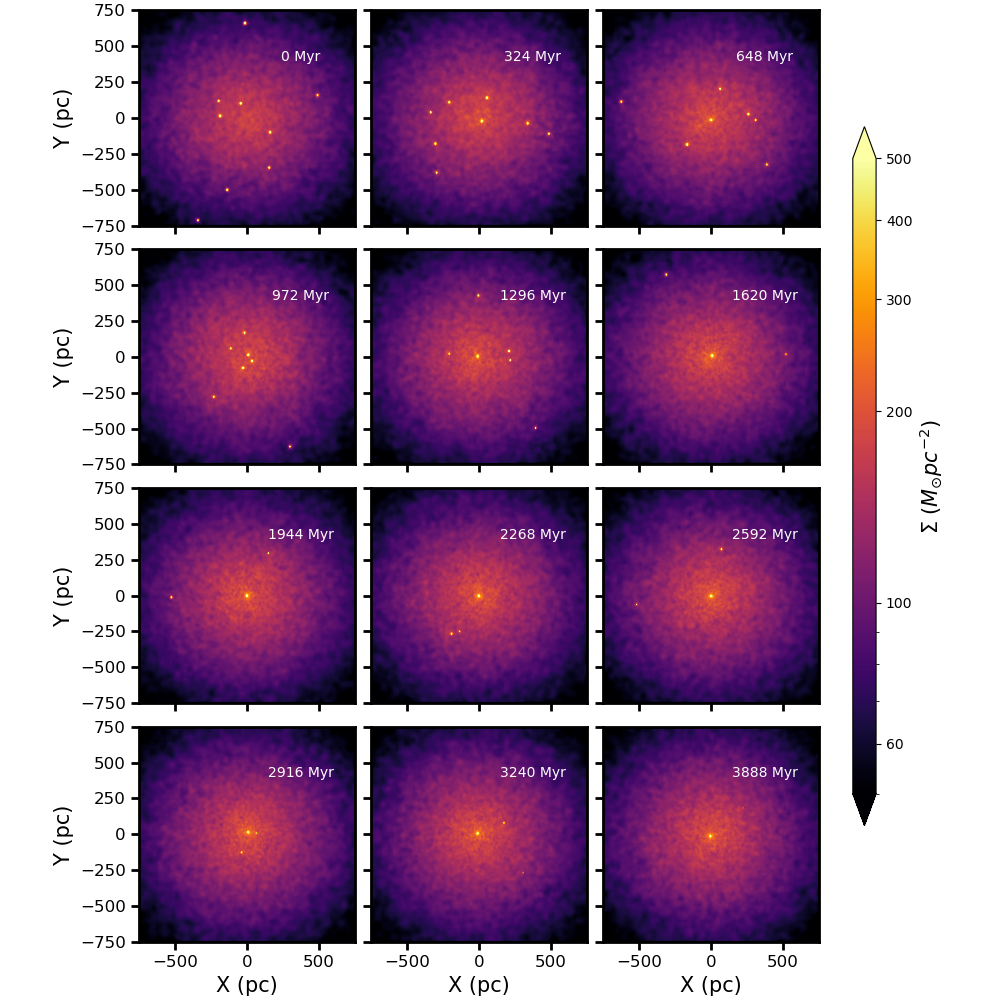}
    \caption{Projected surface density of UGC 7346 and GC system on xy-plane at various times from start till end of our simulation for the case of IC1.}
    \label{fig:res-ic1-snap-projections}
\end{figure*}

We randomly distributed each GC within the GC half-number radius for UGC 7346's observed GC system (taken from JR23), which amounts to $1\, \text{kpc}$. The GCs were placed on sub-circular orbits, with initial velocities set to 50\% of the local circular speed, to represent the eccentric orbital distribution commonly observed in Milky Way GCs \citep{vasiliev2021}.

We created a total of three different initial conditions for the GC system. 
In two of these initial conditions (IC1 and IC2), we gave the GCs different starting positions with respect to the galactic center, in order to check the effect of initial separation on the NSC formation process. 
Table \ref{tab:gc-params} lists the key parameters used for GCs in IC1 and IC2.
In the third initial condition IC3, we kept the same GC starting positions as in IC1 but instead increased the mass of five out of nine GCs by a factor of 10. 
This was done so that we can test how NSC formation would be affected if more massive GCs were involved.
Table \ref{tab:gc-params} lists the key parameters used for GCs in IC3.
The total mass available for NSC formation in IC3 increases to $9.9\times 10^6\, M_{\odot}$.

Once our initial conditions are set, we co-evolve the galaxy and its GC system using direct N-body code $\phi$-GPU \citep{berczik+11,Berczik2013}. $\phi$-GPU employs a direct summation of pairwise forces and integrates particle orbits using a fourth-order Hermite scheme with individual block time steps. For some recent applications of the current version of the $\phi$-GPU code the reader will be redirected to the papers: \cite{LLBCS2012, WBSK2014, HZLBS2018}. In the code, gravitational interactions are calculated using:
\begin{equation}
F_{ij} = -\frac{G m_i*m_j}{\left(r_{ij}^2 + \epsilon_{ij}^2\right)^{1/2}},
\label{eq:plummer}
\end{equation}
where \(r_{ij}\) is the separation between particles \(i\) and \(j\). Each particle is assigned its own softening length, and the effective pairwise softening is defined as:
\begin{equation}
\epsilon_{ij} = \left(\frac{\epsilon_i^2 + \epsilon_j^2}{2}\right)^{1/2}.
\label{eq:softmix}
\end{equation}
We assign a softening length of \(10\,\mathrm{pc}\) to the galaxy particles when evaluating Equation~(\ref{eq:softmix}). 
This choice reflects the role of the galaxy as a smooth background potential that drives GC infall through dynamical friction and promotes disruption through tidal effects.
It also suppresses artificially strong encounters between the relatively massive galaxy particles (\(\sim 500\,M_\odot\)) and the GCs. For GC particles, our adopted softening is $0.1\ \text{pc}$, an order of magnitude smaller than the effective radii of GCs, ensuring that GC internal dynamics is resolved to very small spatial scales.

\subsection{GC stability analysis}\label{subsec:ics_gc_stability}
Due to the limited number of particles used in N-body simulations, a GC cannot be modeled with realistic mass resolution ($\sim 1\, M_{\odot}$ particles) for long time, in our case $\sim$ Gyr time scale.
To determine an appropriate mass resolution to model all GCs with, we first modeled only one GC (GC 5) together with the host galaxy in low, medium, and high mass resolutions.
Mass resolution for each variant ranges from $8\, M_{\odot}$ for high resolution, $25\, M_{\odot}$ for intermediate resolution and $100\, M_{\odot}$ for low resolution.
Each variant was then inserted into the single component galaxy at a distance of $r=500\, \text{pc}$ in the x direction, with a velocity equal to half of their circular velocity $\sqrt{GM(r)/r}$ in the y direction. \par

Top panel of Figure \ref{fig:test_ics_orbits} shows the orbital decay of GC 5 for different mass resolutions. 
We notice that the orbital evolution is effectively independent of mass resolution for our choices. 
The mass loss rate on the other hand shows a noticeable difference (Figure \ref{fig:test_ics_orbits} bottom panel).
The low-resolution realization of GC 5 shows higher mass-loss rates, both within its tidal radius and within $5\,R_e$, than the intermediate- and high-resolution cases. This is likely due to enhanced two-body relaxation in the low-resolution model, which can produce additional numerical heating and thereby accelerate mass loss in the external tidal field. \par
Although the highest mass resolution may present the most accurate results, we chose to model our GCs in medium mass resolution ($25\, M_{\odot}$ particles) since GC stability is largely consistent between medium and high resolution. This allows us to obtain reasonably accurate results without amplifying computational cost (and time) too much.

\section{Results}\label{sec:results}
Here we describe the results of three cases of formation of NSC1, NSC2, and NSC3 from GC initial conditions IC1, IC2, and IC3 respectively.

\begin{figure*}
    \centering
    \includegraphics[width=\linewidth]{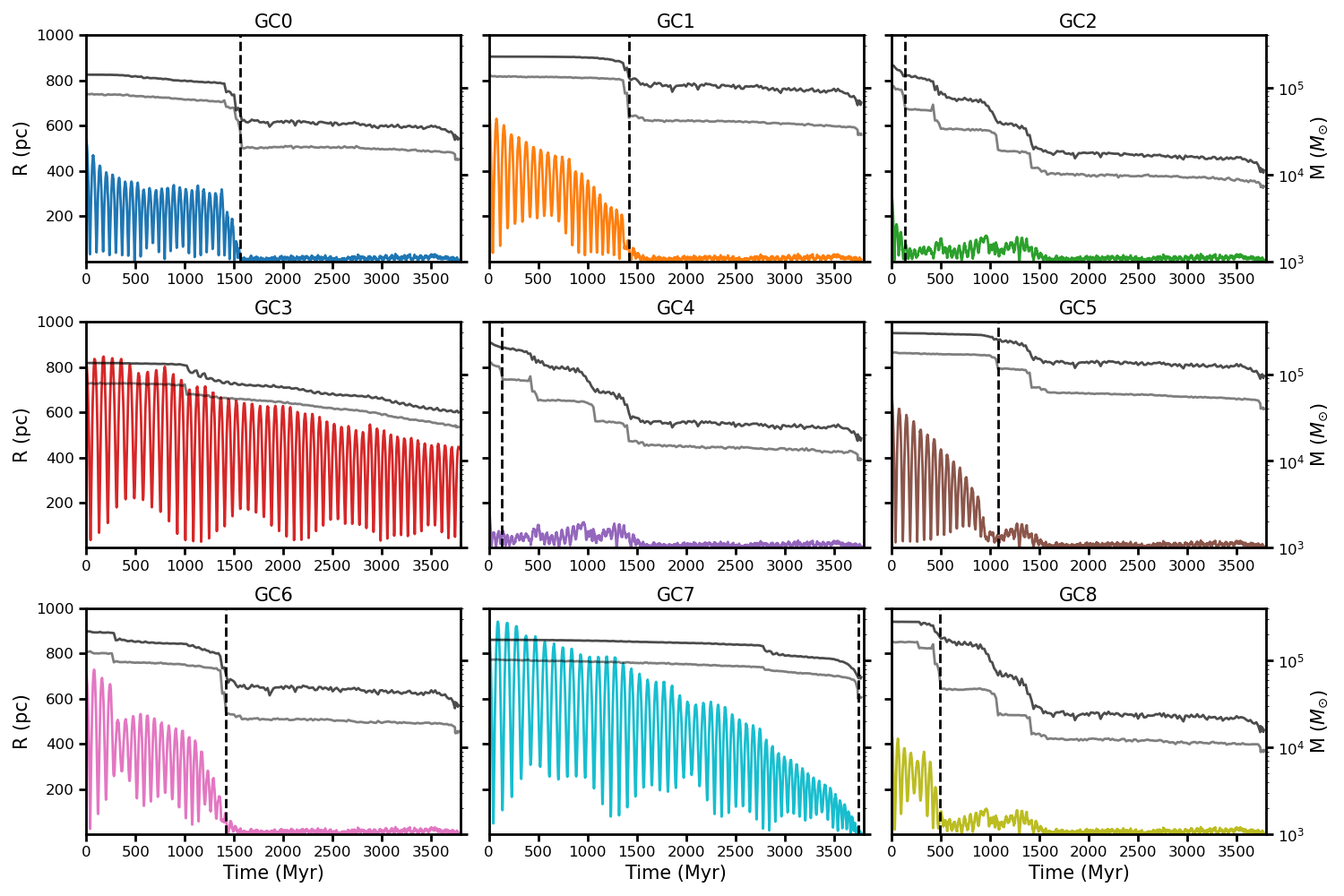}
    \caption{Twin axis plot of orbital decay and mass loss of GCs in IC1. Dark and light gray lines represent total mass within tidal radius and $5R_e$ (equivalent to half-mass radius). Vertical dashed lines represent the time each GC merged at the center.}
    \label{fig:res-ic1-orbits-mass-loss}
\end{figure*}

\subsection{IC1}
The time evolution of GCs system embedded in their host galaxy for IC1 is shown in figure \ref{fig:res-ic1-snap-projections}. The figure shows projected surface density maps for snapshots at various (labeled) times during the run. We notice that all but two remaining clusters have reached the center in about $1.6\, \text{Gyr}$, resulting in a high density environment at the center. 
Figure \ref{fig:res-ic1-orbits-mass-loss} shows the orbital evolution of all GCs relative to the galactic density center as a function of time, along with their tidal mass loss measured within the tidal radius and within $5\,R_e$.
 All GCs sink towards the center because of orbital energy loss caused by dynamical friction against the background stellar distribution of the host galaxy. 

Owing to its low mass and distant starting position, GC 3 was strongly tidally stripped, dynamical friction acting on it weakened, and it failed to reach the center within the simulation time. \par

We define the total mass of a GC at any time as the mass contained within its tidal radius (centered on the density center of the GC).  
The GCs lose mass to the host galaxy at a slow rate until they reach the central $200\, \text{pc}$ at which point we witness a sudden decrease in cluster mass caused by stronger tidal forces \citep{moreno2024influence} before merging with the central cluster. \par

Table~\ref{tab:res-merged} lists the time at which each GC reaches the central region of the galaxy, which we define as when the distance between the density center of the GC and that of the galaxy is less than $1\, \text{pc}$. 
GCs 2 and 4 were the first to reach the center in just over $100\, \text{Myr}$ (represented by the vertical dashed line for each GC) and merge given their smaller initial distances from the center of galaxy. Their merger defined a new high density region at the center that grew with subsequent merger of GCs. Subsequently, GC 8 and GC 5 reached the center.
GCs 0, 1, and 6 merged in quick succession, at a time around $1.5\, \text{Gyr}$. Most of the central cluster mass was assembled by this time. Finally, GC 7 reached and merged with the central cluster at a time of about $3.7\, \text{Gyr}$. \par

\begin{figure}
    \centering
    \includegraphics[width=\linewidth]{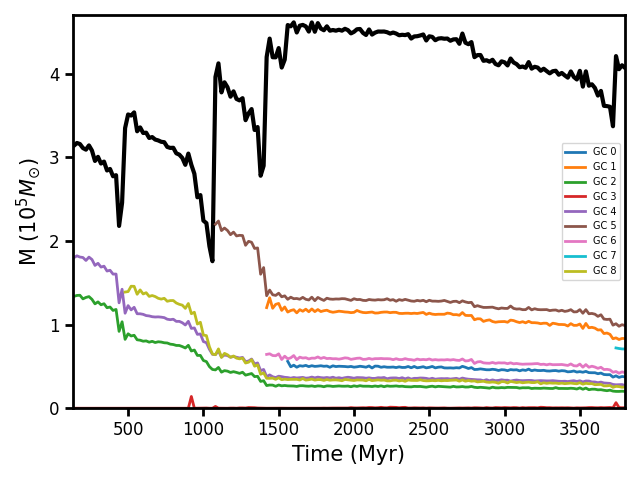}
    \includegraphics[width=\linewidth]{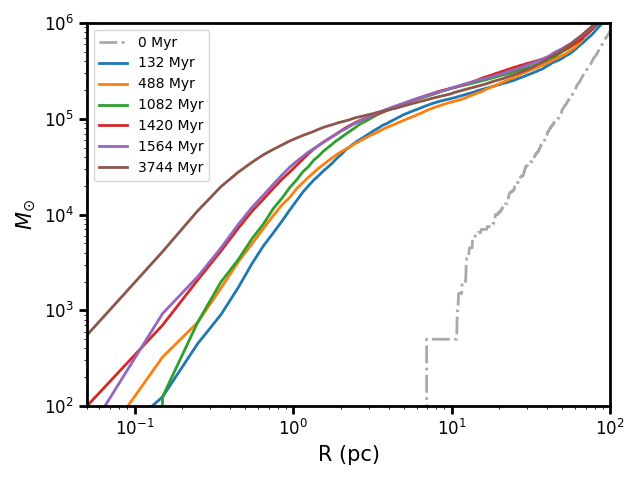}
    \caption{Top panel: Time evolution of NSC1 total mass (solid black line), and individual contribution to the total mass of NSC1 by each GC within $50\, \text{pc}$.
    Bottom panel: Cumulative mass profile of the NSC1 immediately after each GC merger.}
    \label{fig:res-ic1-gc-contribution}
\end{figure}

In the top panel of figure \ref{fig:res-ic1-gc-contribution}, we plot with respect to time how mass accumulates at the center of the galaxy. At any point in time, the solid black line represents the sum of individual GC particle masses within $50\, \text{pc}$ of the density center of the galaxy.
We notice that as a GC approaches the center, some mass from previously settled GCs is pushed out of the $50\, \text{pc}$ region, which is then compensated for by the incoming GC's own mass contribution. 
Despite this, the mass concentration at the center keeps on increasing with each GC merger (bottom panel of figure \ref{fig:res-ic1-gc-contribution}).

As shown in Table \ref{tab:res-merged}, GC 7 has the highest retained mass fraction at the end of the simulation because it reaches the center last, whereas GC 5 contributes the largest mass at merger.

\begin{figure}
    \centering
    \includegraphics[width=\linewidth]{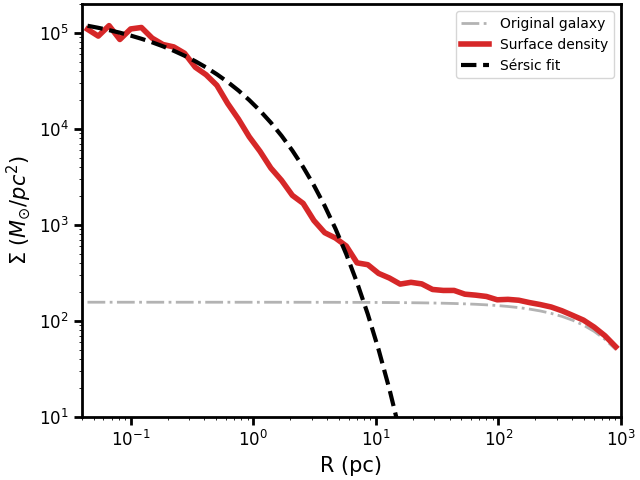}
    \includegraphics[width=\linewidth]{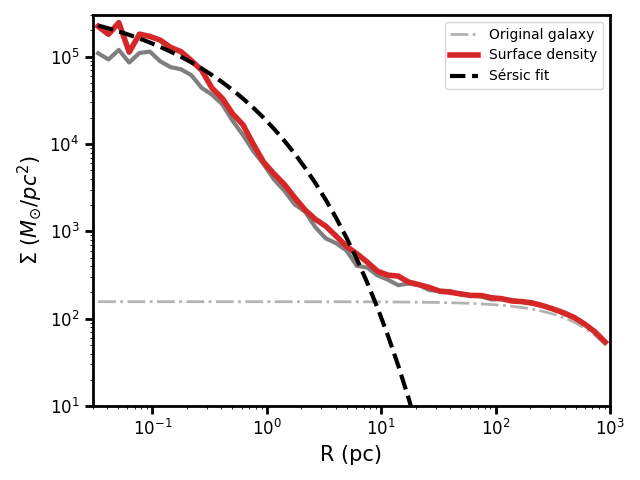}
    \includegraphics[width=\linewidth]{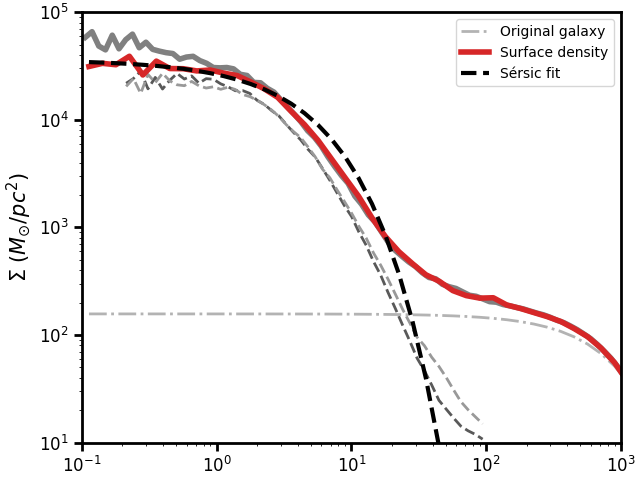}
    \caption{Surface density profile of UGC 7346 + NSC1, NSC2 and NSC3 (from top to bottom respectively). Dash-dotted line shows the surface brightness profile of UGC 7346 at $t=0$. Middle panel: Surface density of NSC1 is shown as a solid grey line for comparison. Bottom panel: Solid grey line represents the surface density of NSC3 at $1.6\, \text{Gyr}$. Thin dashed lines show the surface density of the clusters C1 and C2 observed in IC3.}
    \label{fig:res-nsc-fits-ic1}
\end{figure}

In top panel of Figure \ref{fig:res-nsc-fits-ic1}, we have plotted the surface density of the central cluster that formed in IC1.
We can observe a significant rise in surface density compared to the original galaxy's profile in the inner region smaller than $50\, \text{pc}$, a clear sign that a NSC has formed (NSC1, corresponding to IC1).
We define the total mass contained within this visual extent as mass of NSC1, which amounts to $4.07\times 10^5\, M_{\odot}$ (22\% of the available GC system mass). \par
To determine the NSC's effective radius and S\'ersic index, we employed least-squares curve fitting over the inner surface density profile using the projected S\'ersic density formula as described in \cite{vitral_mamon_2020}:
\begin{equation}\label{eq:projected_sersic_vitral}
    \Sigma(R) = \frac{M}{4\pi R_e^2} \frac{b_n^{2n}}{2n \Gamma (2n)} \ exp \left[ -b_n \left( \frac{R}{R_e} \right) ^ {1/n} \right]
\end{equation}
 Here $\Gamma$ is the complete gamma function. $Re$, $n$ and $b_n$ are the same as in Equation \ref{eq:deprojected_sersic}, and $M$ is fixed to the mass within the visual extent of NSC1.
Table \ref{tab:res-merged} shows the obtained fit parameters.
The NSC is compact with an effective radius of $2\, \text{pc}$, and a S\'ersic index of $n=1.85$. \par

\begin{table*}

\centering
\begin{tabular}{@{}ccccccc@{}}
\toprule
ID/quantity
  & $T_{\mathrm{NSC1}}$ 
  & $T_{\mathrm{NSC2}}$ 
  & $T_{\mathrm{NSC3}}$
  & $M_{\mathrm{NSC1}}$ 
  & $M_{\mathrm{NSC2}}$ 
  & $M_{\mathrm{NSC3}}$ \\
  & (Myr) & (Myr) & (Myr) & (\%) & (\%) & (\%) \\
\midrule
GC 0 & 1564 & 3474 & 346 & 18.3 & 12.1 & 35.4 \\
GC 1 & 1420 & 2312 & 106 & 29.7 & 40.5 & 50.8 \\
GC 2 & 132  & 2874 & --  & 6.2  & 23.3 & --   \\
GC 3 & --   & 1314 & 106 & --   & 4.0  & 20.6 \\
GC 4 & 132  & 798  & --  & 7.9  & 11.2 & --   \\
GC 5 & 1082 & 798  & --  & 32.1 & 19.3 & --   \\
GC 6 & 1420 & 1038 & 60  & 14.1 & 15.4 & 56.9 \\
GC 7 & 3744 & 2862 & 60  & 37.0 & 28.1 & 43.8 \\
GC 8 & 488  & 1542 & --  & 5.6  & 23.7 & --   \\
\midrule
\multicolumn{7}{c}{NSC structural parameters} \\
\midrule
Fit parameter 
  & \multicolumn{2}{c}{NSC1}
  & \multicolumn{2}{c}{NSC2} 
  & \multicolumn{2}{c}{NSC3} 
  \\
S\'ersic index $n$ 
  & \multicolumn{2}{c}{1.85} 
  & \multicolumn{2}{c}{2.47} 
  & \multicolumn{2}{c}{1.16} 
  \\
Effective radius $R_e$ (pc) 
  & \multicolumn{2}{c}{2.03} 
  & \multicolumn{2}{c}{2.17} 
  & \multicolumn{2}{c}{8.58} 
  \\
Total mass ($M_{\odot}$) 
  & \multicolumn{2}{c}{$4.07 \times 10^5$} 
  & \multicolumn{2}{c}{$4.50 \times 10^5$} 
  & \multicolumn{2}{c}{$4.61 \times 10^6$} 
  \\
\bottomrule
\end{tabular}
\caption{Columns 2--4: merger time (in Myr) of each GC at the center for each initial condition. 
Columns 5--7: percentage of the total original mass retained/contributed by each GC at the end of the simulations. 
The lower block lists the best-fit S\'ersic parameters for the NSC in each case, with the NSC mass fixed
to the mass within its visual extent.}
\label{tab:res-merged}
\end{table*}

\subsection{IC2}

\begin{figure*}
    \centering
    \includegraphics[width=\linewidth]{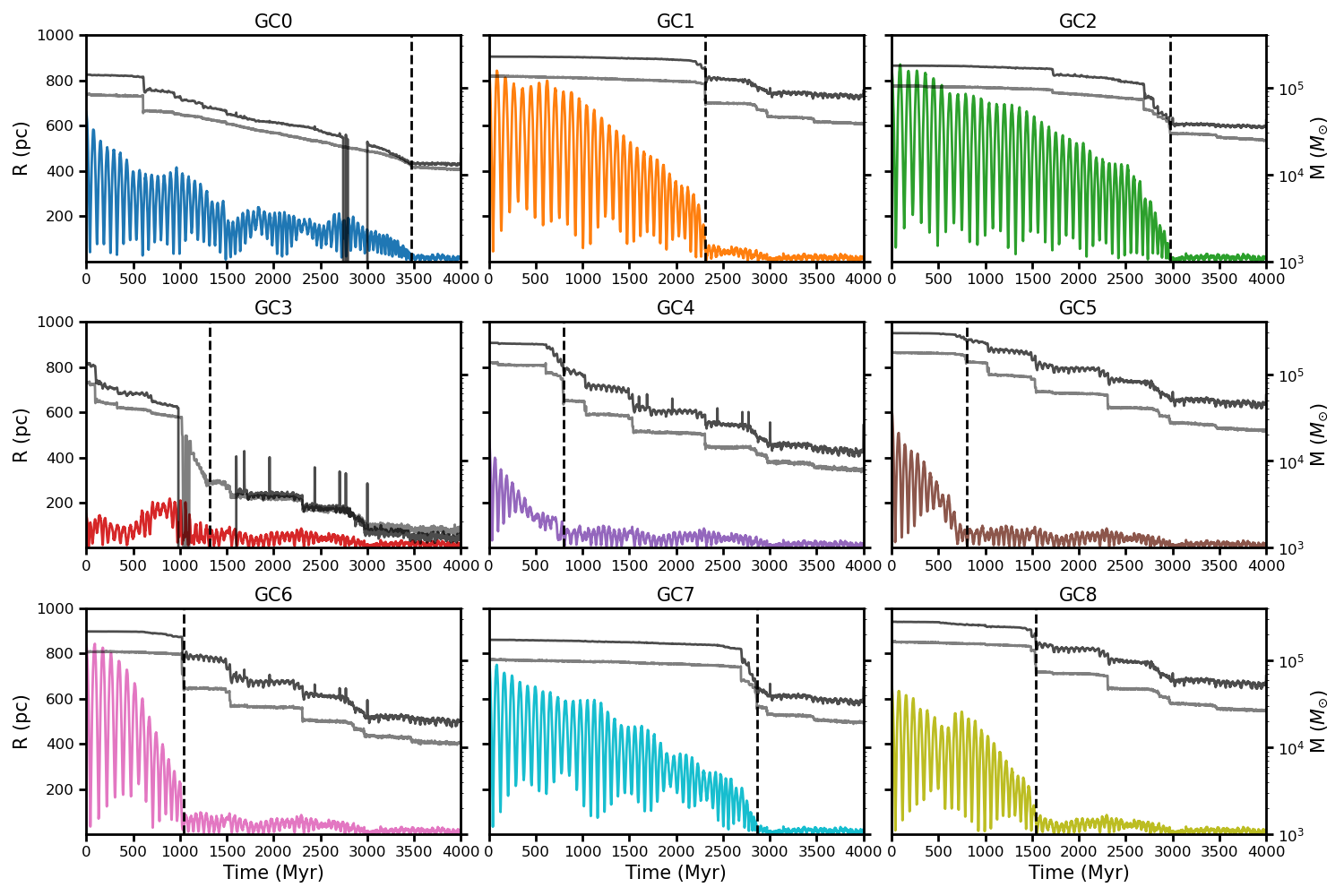}
    \caption{Twin axis plot of orbital decay and mass loss of GCs in IC2. Dark and light gray lines represent total mass within tidal radius and $5R_e$ (equivalent to half-mass radius). Vertical dashed lines represent the time each GC merged at the center.}
    \label{fig:res-ic2-orbits-mass-loss}
\end{figure*}
\begin{figure}
    \centering
    \includegraphics[width=\linewidth]{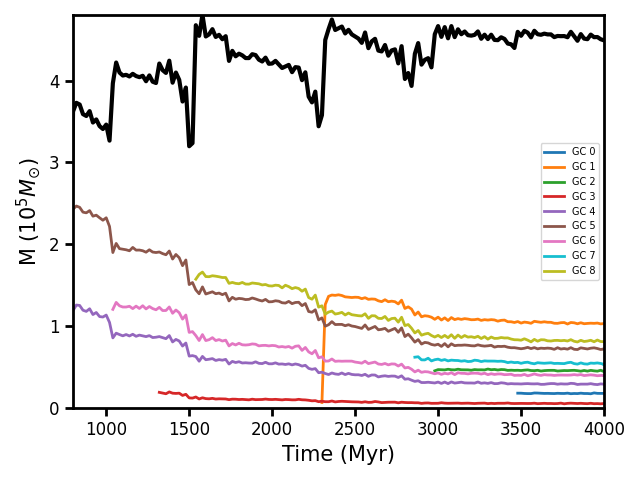}
    \includegraphics[width=\linewidth]{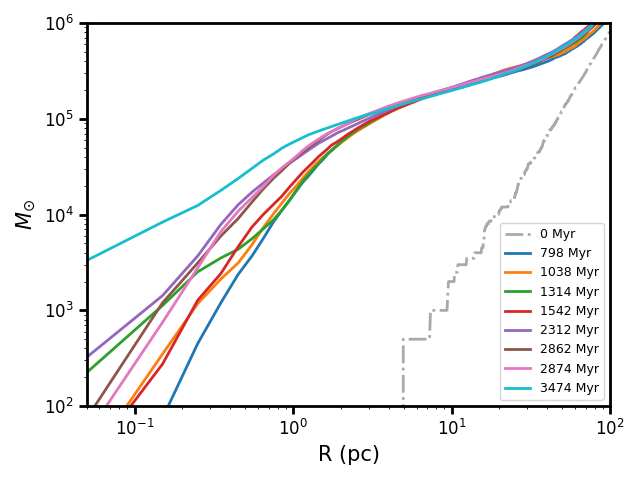}
    \caption{Top panel: Time evolution of NSC2 total mass (black solid line), and individual contribution to the total mass of NSC2 by each GC within $50\, \text{pc}$.
    Bottom panel: Cumulative mass profile of the NSC2 immediately after each GC merger.}
    \label{fig:res-ic2-gc-contribution}
\end{figure}

The GCs in IC2 have different starting positions compared to IC1. Here, GCs 4 and 5 were the first to merge at the center around $798\, \text{Myr}$ (Figure \ref{fig:res-ic2-orbits-mass-loss}), consequently defining the new density center of the galaxy. 
All GCs managed to reach the center within $3.5\, \text{Gyr}$. 
However, as evident from mass loss in Figure \ref{fig:res-ic2-orbits-mass-loss}, GC 3 lost a large part of its mass early on and possessed mass of the order of $10^3$ $M_{\odot}$ by the time it merged at the center.
Similar case is for GC 0, which had an early collision with GC 4 around $600\, \text{Myr}$ (evident in first panel of Figure \ref{fig:res-ic2-orbits-mass-loss} by the sharp drop in mass just after $500\, \text{Myr}$) that triggered accelerated mass loss. \par

Figure \ref{fig:res-ic2-gc-contribution} shows how the central cluster grows over time. Given that the most massive GC reached the center first, the subsequent growth of the central cluster is less drastic in comparison to NSC1. 
The surface density profile (middle panel of figure \ref{fig:res-nsc-fits-ic1}) shows a distinct over-density at distances less than $50\, \text{pc}$, which qualifies the central cluster as NSC2. In this case $M_{nsc}=4.49\times 10^5$ $M_{\odot}$.
NSC2 has an effective radius similar to that of NSC1, but possesses a higher S\'ersic index of $n=2.47$.

\subsection{IC3}
Here we explore a fiducial scenario in which five of our clusters are 10 times more massive than those in the previous two cases, with masses of the order of $\sim 10^6$ $M_{\odot}$. The GCs in IC3 are initiated with the same starting positions and velocities as in IC1. 
Understandably, all massive GCs reached the center earlier—within $400\, \text{Myr}$—owing to the stronger dynamical friction they experienced as a result of their higher masses. 
The inspiralling GCs first formed a double cluster in roughly $100\, \text{Myr}$ where GCs 6 and 7 merged to form cluster C1 at $60\, \text{Myr}$ and GCs 1 and 3 merged to form cluster C2 at $106\, \text{Myr}$.

GC 0 merged with C2 at $346\, \text{Myr}$, thereby becoming the more massive of the two clusters.
Both clusters have a combined mass of $4.92\times 10^6\, M_{\odot}$. \par

\begin{figure*}
    \centering
    \includegraphics[width=0.9\linewidth]{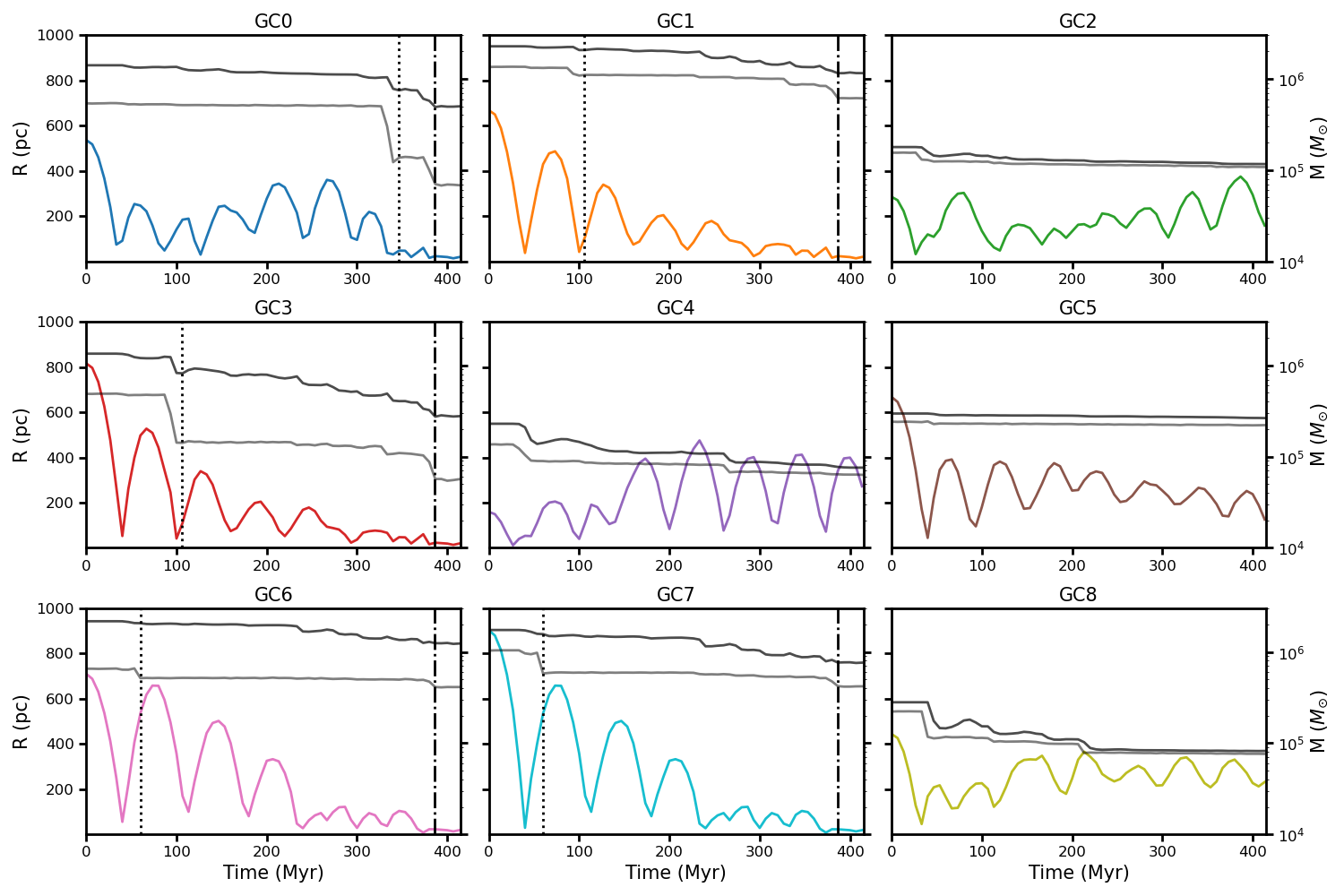}
    \caption{Twin axis plot of orbital decay and mass loss of GCs in IC3. Dark and light gray lines represent total mass within tidal radius and $5\, R_e$ (equivalent to half-mass radius). Dash-dotted line indicates merger time of C1 and C2, whereas dotted lines indicate time when each GC merged with their respective cluster.}
    \label{fig:res-ic3-orbits}
\end{figure*}

Clusters C1 and C2 eventually merged at $386\, \text{Myr}$ to form a single massive central cluster whose surface density profile is shown in bottom panel of figure \ref{fig:res-nsc-fits-ic1}. Clearly, this central cluster can also be classified as an NSC. Subsequently, GCs 5 and 2 managed to reach the center at $1120\, \text{Myr}$ and $1615\, \text{Myr}$ respectively. \par
 
NSC3 possesses a total mass of $4.79\times 10^6\, M_{\odot}$ within its visual extent of $100\, \text{pc}$, greater than the combined mass of NSC1 and NSC2. This suggests that more massive GCs efficiently add mass to the center, as they resist tidal mass loss due to their stronger internal gravity and also move to the center more rapidly because they experience a stronger dynamical friction force. Compared to NSC1 and NSC2, NSC3 is more extended and has lower $\Sigma$ values toward the center.

\section{Discussions}
It is common to analyze the correlations between the structural parameters of NSCs and their hosts \citep{walcher_etal_2005, cote_etal_2006_acsvcs_viii, erwin_gadotti_2012, georgiev_etal_2016, spengler_etal_2017, ordenes_briceno_2018_ngfs_iv, sanchez_janssen_2019_ngvs_xxiii, r_pechetti_etal_2020, nils_hoyer_etal_2023}. Here, we examine several key parameters of NSCs formed in the three cases considered and compare them with established correlations in the literature. 

\subsection{NSC mass}
\begin{figure}
    \centering
    \includegraphics[width=\linewidth]{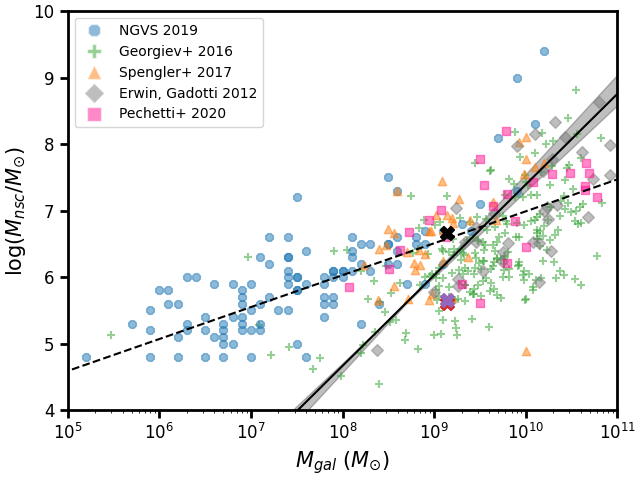}
    \caption{Comparison of our obtained NSC masses with NSCs in literature. The sample from \cite{sanchez_janssen_2019_ngvs_xxiii} and \cite{spengler_etal_2017} represent Virgo Cluster NSCs. The samples from \cite{georgiev_etal_2016}, \cite{erwin_gadotti_2012} represent late type field spirals, and \cite{r_pechetti_etal_2020} represents local volume galaxies. The dashed black line represents the $M_{gal}-M_{nsc}$ scaling relation presented by \cite{neumayer_etal_2020} and the solid black line shows the scaling relation for early type galaxies by \citet{georgiev_etal_2016}. The red and purple crosses represent NSC1 and NSC2, and the black cross represents the NSC3.}
    \label{fig:res-surveys}
\end{figure}

Table \ref{tab:res-merged} shows the masses of the NSCs formed during the inspiral of GCs for three different cases that we studied. There is a rich literature suggesting that the mass of a NSC and its host mass are well correlated \citep{Graham2003,Scott2013,georgiev_boker_2014,sanchez_janssen_2019_ngvs_xxiii}. We plotted $M_{nsc}-M_{gal}$ relation  in Figure \ref{fig:res-surveys}. For our case, both NSC1 and NSC2 are undermassive compared to the predicted $\log(M_{nsc}/M_{\odot})$ values from \citet{neumayer_etal_2020}'s. They are more in line with the NSC scaling relation for early-type galaxies of \citet{georgiev_etal_2016}, which resonates well with UGC 7346 being an early-type dwarf galaxy. 
NSC3 matches the predicted value from \citet{neumayer_etal_2020} quite nicely, owing to the more massive GCs.
The mean $\log(M_{nsc}/M_{\odot})$ for the shown Virgo Cluster sample is 6.37, with a standard deviation of 0.84. 
All three of our obtained NSCs are well within $1\sigma$ of the shown sample.

We argue that the NSC masses presented here may represent a lower limit on the potential NSC that can form in UGC 7346. 
Our current understanding of NSC formation is that GC collapse and in-situ star formation can work in conjunction to form a NSC, especially for a host galaxy of $10^9\, M_{\odot}$ \citep{neumayer_etal_2020,fahrion_etal_2021}. 
However, it is difficult to quantify how much star formation will contribute to building the NSC for our case.
\cite{fahrion2022} derived the fraction of NSC mass contributed by star formation $(f_{in})$ for 119  nucleated galaxies based on their observed NSC masses and GC system masses. They found that $f_{in}$ is strongly correlated with NSC mass, and also with the ratio $M_{nsc}/M_{GCS}$. Lower mass NSCs and galaxies where $M_{GCS}>M_{nsc}$ possess a low $f_{in}$ (meaning they predominantly formed through GC collapse), and vice versa. \par

JR23’s photometric analysis determined that UGC 7346 is a blue-cored dE, with bluer color in its central region and redder color in its outer regions. 
\cite{urich2017young} and \cite{lisker2006virgo} have observed such galaxies to possess younger star populations in their center, implying recent or ongoing star formation. Therefore, the potential impact of star formation on the NSC cannot be ruled out for the case for UGC 7346.


\subsection{NSC S\'ersic index and surface density profile}
S\'ersic index of the NSC defines the degree of central concentration of stars. Nuclei with high $n$ tend to also be brighter \citep{spengler_etal_2017}.
The S\'ersic indices of our NSCs possess a certain scatter (see table \ref{tab:res-merged}) and do not show a clear pattern. 
NSC1 and NSC2 possess similar mass and yet their S\'ersic indices differ.
NSC3 despite being the most massive NSC in our simulations possesses the smallest S\'ersic index. 
The lack of a clear pattern in our NSC S\'ersic indices is consistent with the observations of \citet{r_pechetti_etal_2020} and \citet{ashok2023composite}, who observed indices ranging from $n<1$ to $n=6$  or higher. These studies found a median S\'ersic index value of 1.9 and 2.7 respectively, the former agreeing well with the S\'ersic index of NSC1 (although it should be noted that the median value of Pechetti et al. is after excluding NSCs with $n>6$). \par 

In regards to fitting the surface brightness profiles of the NSCs, we note that our fits deviate from the observed profile in the middle regions.
This is especially true for NSC1 and NSC2, where beyond $1\, \text{pc}$ the fit exceeds the observed brightness before falling to obscurity.
One could improve the quality of the fitting by using two S\'ersic components, however, doing so leads to degenerate components (many different combinations of $n$ and $R_e$ that fit to the same surface brightness profile).
Therefore, we choose to keep the shown single component fits. \par
 
\subsection{NSC effective radius}

\begin{figure}
    \centering
    \includegraphics[width=\linewidth]{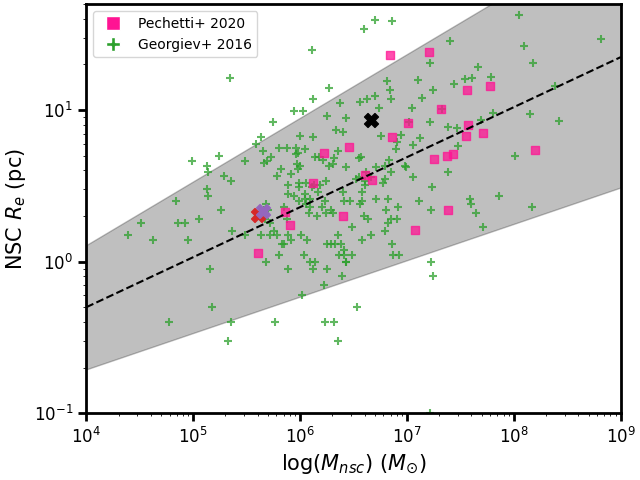}
    \includegraphics[width=\linewidth]{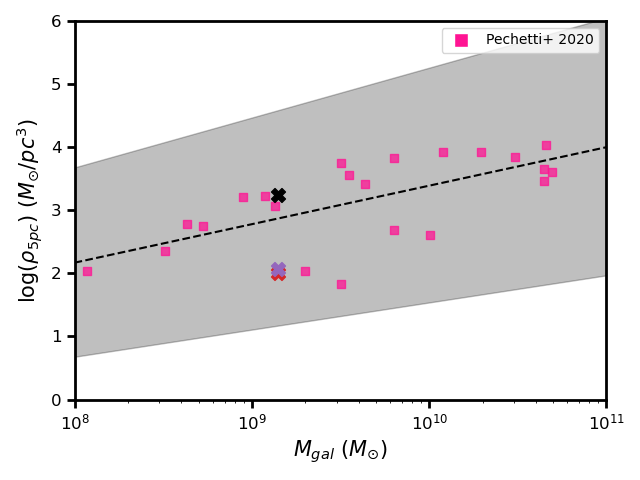}
    \caption{Top panel: Comparison of our obtained NSC effective radii with NSCs in literature. The solid black line represents the $M_{nsc}-R_{e, nsc}$ scaling relation presented in \cite{r_pechetti_etal_2020} along with confidence interval. Bottom panel: Comparison of volume density of our NSCs with NSCs in literature. Solid black line represents the $\rho_{5pc,nsc}-M_{gal}$ scaling relation presented in \cite{r_pechetti_etal_2020} along with confidence interval.}
    \label{fig:res-Mnsc-Re}
\end{figure}

\begin{figure}
    \centering
    \includegraphics[width=\linewidth]{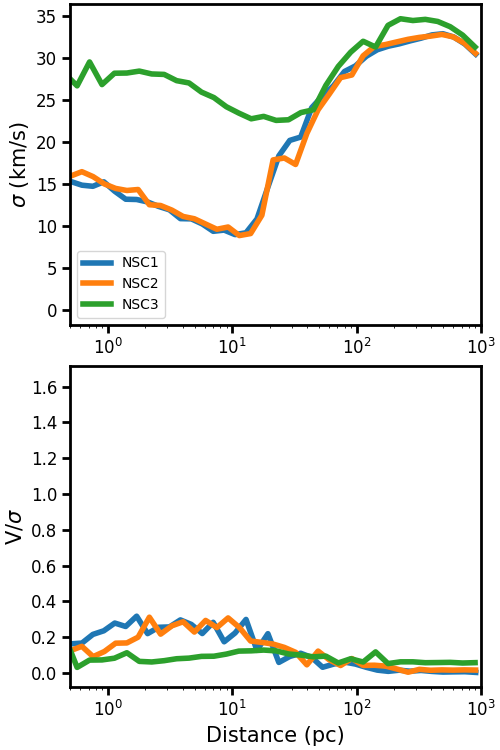}
    \caption{Velocity dispersion (top panel) and $v/\sigma$ (bottom panel) profiles of all the NSCs and galaxy as a function of distance from the center.}
    \label{fig:res-nsc-kinematics}
\end{figure}

Figure \ref{fig:res-Mnsc-Re} (top panel) depicts the correlation between NSC mass and effective radius for the data obtained from the literature together with those obtained for our case (see table 5). More massive NSCs are also larger in size \citep{r_pechetti_etal_2020, georgiev_etal_2016}, a fact that our results are consistent with. 
NSC3 is more massive than NSC1 and NSC2, and also possesses a larger effective radius.
The mean $\log(R_e)$ for the sample shown in Figure \ref{fig:res-Mnsc-Re} is 0.53 (equal to $3.39\, \text{pc}$), with a standard deviation of 0.43 ($\pm \ 2.7\, \text{pc}$). NSC1 and NSC2 lie within $1\sigma$ of the sample distribution, whereas NSC3 falls within $2\sigma$.
\subsection{NSC volume density}

NSC and massive black hole (MBH) have been shown to coexist in the center of galaxies \citep{ngu17,ngu19,ask23}. Several interesting astrophysical phenomena, such as tidal disruption events (TDEs), merger of black holes \citep{Khan_Holley-Bockelmann2021,khan2025,KHB2025} and gravitational wave emission \citep{LISA2023LRR,colpi24} are attributed to the presence of MBH in the center of galaxies. These phenomena are directly linked to the central density of stars in galactic nuclei. For instance, the fraction of stars on orbits that interact with central MBH \citep{Baile15} is directly related to the stellar density within the sphere of influence of MBH \citep{khan+12a,khan18a}. Also, the TDE rate is expected to be higher for dense stellar environments.

\citet{r_pechetti_etal_2020} calculated the deprojected density profiles for their sample of 29 galaxies, and noted that higher mass galaxies host more dense NSCs.
We calculate the NSC stellar density at $5 ~\text{pc}$ ($\rho_{5\text{pc}}$) from the center and compare it with the observed values of \citet{r_pechetti_etal_2020}, who quantified this trend through the $M_{\text{gal}}$–$\log(\rho_{5\text{pc},\text{nsc}})$ scaling relation (their Equation 12), shown in the bottom panel of Figure~\ref{fig:res-Mnsc-Re}.
The shown sample has a mean $\log(\rho_{5pc,nsc})$ of 2.39, with a standard deviation of 1.47. Again, all our NSCs fit within the $1\sigma$ limit.

\pagebreak
\subsection{NSC Kinematics}
Figure \ref{fig:res-nsc-kinematics} shows the kinematic properties of the NSCs formed in our simulation suite. Overall, the NSCs exhibit higher velocity dispersion towards the center, particularly within the inner $10\,\text{pc}$, and low rotation, indicating that they are predominantly dispersion-supported systems. While all three NSCs share this general characteristic, NSC3 stands out with higher velocity dispersion and virtually no rotation. 

\section{Summary \& Conclusion}
we performed N-body simulations to study the evolution of UGC 7346 and nine of its observed GCs. In two cases, we adopted masses inferred from the clusters’ light profiles and colors. In a third, fiducial case, some of the clusters were assumed to be ten times more massive than those in the first two cases. For case 1 and 2, we varied the initial positions and orbits of GCs to check the robustness of our findings. Below, we summarize key findings of this study. 

\begin{itemize}
    \item For cases IC1 and IC2, representative of the GC masses in UGC~7346, most clusters reach the galaxy center within \(\sim 2\text{--}3\,\mathrm{Gyr}\), experiencing varying degrees of mass loss depending on their initial masses and orbital parameters.

    \item For all cases, we notice an over-density by witnessing a rise in surface density profile as we move inwards from $\sim 20\, \text{pc}$. This is a typical observational signature used to infer the presence of an NSC in galaxies.

    \item The NSC masses obtained for the first two cases, where we used the inferred GC masses, are approximately $4.1 \times 10^5\,M_{\odot}$ and $4.5 \times 10^5\, M_{\odot}$. These values fall within the observed range of NSC masses, although they are slightly on the lower side compared to the average NSC–galaxy mass correlation. In the case where we used more massive GCs, the resulting NSC mass is more in line with the expected value for a galaxy of this mass.

    \item The inspiral and NSC formation timescale is on the order of a Gyr for the typical GC masses adopted in this study, within $1\, \text{kpc}$ — a scale comparable to the effective radius of dwarf galaxies.

    \item For the case of more massive GCs the NSC formation time is less than a Gyr. We notice that a double nucleus separated by roughly $200\,  \text{pc}$ forms and survives for $280\, \text{Myr}$ which may be confused with double nucleus created by merger of galaxies. 

    \item The remaining properties of all NSCs formed in our study, such as effective radius, Sérsic index, and central density, fall well within the observed range of NSC parameters.
\end{itemize}

In conclusion, our results show that GCs inspiral is a viable and efficient channel for NSC formation in reasonable timescales. UGC 7346 will have a nuclear star cluster in about a billion year form now as do most of the galaxies in this mass range ($\sim 10^9\, M_{\odot}$). The mass of the NSC formed by dissipation-less channel alone, as is the case in our study, can be regarded as a lower limit. The in situ star formation can add more mass to the center and hence to NSC making it more massive. Another important aspect missing from this study is the presence of a central IMBH and its impact on the formation and evolution of NSC. The results of previous $N$-body simulations of dissipationless NSC formation indicate that a central MBH with $M_{\rm BH} < 10^6 ~M_{\odot}$ can disrupt NSC formation in proportion to its mass \citep{antonini_etal_2012, antonini2013origin, arca_sedda_etal_2015_henize_2_10}. Smaller MBHs do not significantly affect the internal structure of NSCs during formation \citep{antonini2013origin}. However, if a multiple-IMBHs system forms through GCs that deposit several IMBHs in the center, it can potentially partially or completely disrupt the NSC \citep{Khan_Holley-Bockelmann2021,Khan2025BHs}.

Finally, we did not include dark matter in our $N$-body galaxy model. A dwarf galaxy with a stellar mass of $10^{9}\,M_{\odot}$ is typically hosted by a dark matter halo of mass $\sim 10^{11} ~M_{\odot}$. Hydrodynamical simulations (e.g.~\citep{DiCintio2014,Tollet2016,Read2016}) and observations of nearby dwarf galaxies \citep{Oh2015} suggest that halos in the mass range $M_{200} \sim 10^{10}$--$10^{11}\,M_{\odot}$ can develop cores of order $\sim 1~\mathrm{kpc}$. The estimated enclosed dark-matter mass within $1~\mathrm{kpc}$ turns out to be $\sim 10^{8} M_{\odot}$, indicating that dark matter is expected to be subdominant in the inner kiloparsec. The presence of dark matter, depending on its relative contribution compared to stellar mass, would accelerate the orbital decay of infalling GCs.

\section{Acknowledgments}
The authors acknowledge the Space and Astrophysics Research Lab (SARL) and the National Center for GIS and Space Applications (NCGSA), located in the Institute of Space Technology Islamabad. This material is based on work supported by Tamkeen under the NYU Abu Dhabi Research Institute grant CASS. PB thanks the support from the special program of the Polish Academy of Sciences and the U.S. National Academy of Sciences under the Long-term program to support Ukrainian research teams grant No.~PAN.BFB.S.BWZ.329.022.2023. PB appreciate the Polish high-performance computing infrastructure PLGrid (HPC Centre: ACK Cyfronet AGH -- docs.hpc.cyfronet.pl) for providing Helios computer facilities and support within computational grant No.~PLG/2026/019243. PB also acknowledge the Gauss Centre for Supercomputing e.V. (www.gauss-centre.eu) for funding this project by providing computing time through the John von Neumann Institute for Computing (NIC) on the GCS Supercomputer JUWELS Booster and JUPITER Booster at Julich Supercomputing Centre (JSC), Germany. \par 
This research has used NASA's Astrophysics Data System. This research was carried out on the high-performance computing resources at SARL-NCGSA, ACCRE Vanderbilt University, and New York University Abu Dhabi. 

\section{Data Availability Statement}
The data underlying this article will be shared on reasonable request to the corresponding author.

\clearpage

\bibliography{bib_1}{}
\bibliographystyle{aasjournal}



\end{document}